\documentclass{article}

\usepackage{arxiv}

\usepackage[utf8]{inputenc} 
\usepackage[T1]{fontenc}    
\usepackage{hyperref}       
\usepackage{url}            
\usepackage{booktabs}       
\usepackage{amsfonts}       
\usepackage{nicefrac}       
\usepackage{microtype}      
      
\usepackage{lipsum}         
\usepackage{graphicx}
\usepackage{natbib}
\usepackage{doi}
\usepackage{graphicx}
\usepackage{amsmath}
\usepackage{cleveref} 
\usepackage{amssymb}
\usepackage{siunitx}
\usepackage{xparse} 
\usepackage{longtable}
\usepackage{textcomp}
\usepackage{array}
\usepackage{rotating} 
\usepackage{tabularx} 
\usepackage{cite}     
\usepackage{multirow} 
\usepackage{ragged2e}
\usepackage{makecell}
\usepackage{pdflscape}
\usepackage{calc}
\usepackage{supertabular}

\title{Measuring Agile Agreement: Development and Validation of the Manifesto and Principle Scales}

\newif\ifuniqueAffiliation

\ifuniqueAffiliation 
\author{ \href{https://orcid.org/0000-0002-8479-7250}{\includegraphics[scale=0.06]{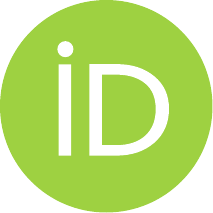}\hspace{1mm}Nicolas Matton}\thanks{Corresponding Author} \\
    University of Namur\\ Faculty of Computer Science\\ rue de Bruxelles 55\\ 5000 Namur, Belgium \\
	\texttt{nicolas.matton@unamur.be} \\
	\And
	\href{https://orcid.org/0000-0002-1816-5685}{\includegraphics[scale=0.06]{orcid.pdf}\hspace{1mm}Anthony~Simonofski} \\
    University of Namur\\ School of Management\\ rue de Bruxelles 55\\ 5000 Namur, Belgium \\
	\texttt{anthony.simonofski@unamur.be} \\
	\And
	\href{https://orcid.org/0000-0002-5488-6383}{\includegraphics[scale=0.06]{orcid.pdf}\hspace{1mm}Marie-Ange Remiche} \\
    University of Namur\\ Faculty of Computer Science\\ rue de Bruxelles 55\\ 5000 Namur, Belgium \\
	\texttt{marie-ange.remiche@unamur.be} \\
	\And
	\href{https://orcid.org/0000-0001-9752-0085}{\includegraphics[scale=0.06]{orcid.pdf}\hspace{1mm}Benoît Vanderose} \\
    University of Namur\\ Faculty of Computer Science\\ rue de Bruxelles 55\\ 5000 Namur, Belgium \\
	\texttt{benoit.vanderose@unamur.be} \\
}
\else
\usepackage{authblk}

\newbox{\orcid}\sbox{\orcid}{\includegraphics[scale=0.06]{orcid.pdf}} 
\author[1]{%
	\href{https://orcid.org/0000-0002-8479-7250}{\usebox{\orcid}\hspace{1mm}Nicolas~Matton\thanks{\texttt{Corresponding Author : nicolas.matton@unamur.be}}}%
}
\author[2]{%
	\href{https://orcid.org/0000-0002-1816-5685}{\usebox{\orcid}\hspace{1mm}Anthony~Simonofski}%
}
\author[1]{%
	\href{https://orcid.org/0000-0002-5488-6383}{\usebox{\orcid}\hspace{1mm}Marie-Ange Remiche}%
}
\author[1]{%
	\href{https://orcid.org/0000-0001-9752-0085}{\usebox{\orcid}\hspace{1mm}Benoît Vanderose}%
}
\affil[1]{University of Namur - Faculty of Computer Science, rue de Bruxelles 55, 5000 Namur, Belgium}
\affil[2]{University of Namur - School of Management, rue de Bruxelles 55, 5000 Namur, Belgium}
\fi

\renewcommand{\headeright}{}
\renewcommand{\undertitle}{}
\renewcommand{\shorttitle}{Measuring Agile Agreement: MAS and PAS instruments}

\hypersetup{
pdftitle={Measuring Agile Agreement: Development and Validation of the Manifesto and Principle Scales},
pdfsubject={q-bio.NC, q-bio.QM},
pdfauthor={Nicolas~Matton, Anthony~Simonofski, Marie-Ange~Remiche, Benoît~Vanderose},
pdfkeywords={Agile Agreement, Agile Values vs. Practices, Agile Software Development, Human Factors},
}

\begin{document}
\maketitle

\begin{abstract}
While the importance of human factors in agile software development is widely acknowledged, the measurement of an individual's "agile agreement" remains an ill-defined and challenging area. A key limitation in existing research is the failure to distinguish between agreement with the abstract, high-level values of the Agile Manifesto and agreement with the concrete, day-to-day practices derived from the 12 Principles.
This paper addresses this methodological gap by presenting the design and validation of two distinct instruments: the novel Manifesto Agreement Scale (MAS), and the Principle Agreement Scale (PAS), which is a systematic adaptation and refinement of a prior instrument.

We detail the systematic process of item creation and selection, survey design, and validation. The results demonstrate that both scales possess important internal consistency and construct validity. A convergence and divergence analysis, including Proportional Odds Logistic Regression, a Bland-Altman plot, and an Intraclass Correlation Coefficient (ICC), reveals that while the two scales are moderately correlated, they are not interchangeable and capture distinct dimensions of agile agreement. The primary contribution of this work is a pair of publicly available instruments, validated within a specific demographic of Belgian IT professionals. These scales represent a critical initial step toward facilitating a more nuanced measurement of agile agreement, distinguishing agile agreement across various levels of perception and aiding in a more refined interpretation of person-agile fit.
\end{abstract}


\section{Introduction}

The sustained adoption of agile methodologies in software development underscores the critical role of human factors in project success \citep{daraojimbaCOMPREHENSIVEREVIEWAGILE2024,gandomaniHowHumanAspects2014}. While the Agile Manifesto famously prioritizes "Individuals and interactions over processes and tools" \citep{fowler2001agile}, the ability to rigorously measure an individual's alignment with this approach remains a significant challenge. The concept of "agile agreement" is frequently invoked, but it often remains an ill-defined and poorly measured construct, hindering evidence-based research and practice.

A fundamental ambiguity pervades existing literature: the distinction between an individual's alignment with abstract \textit{values} versus their agreement with concrete, daily \textit{practices}. This distinction is not merely semantic; it is crucial for understanding how agile is truly adopted and enacted. While numerous instruments assess related constructs like team climate \citep{dutraTACTInsTrumentAssess2022}, global maturity \citep{grenProspectsQuantitativeMeasurement2015}, or the perceived implementation \citep{soPerceptiveAgileMeasurement2009} and relevance \citep{vishnubhotlaUnderstandingPerceivedRelevance2021} of practices, they do not capture this specific individual-level dichotomy. This critical gap forces reliance on proxy~\citep{coelhoDesignModelApplication2020} or aggregated~\citep{stettinaFiveAgileFactors2011} measures, fundamentally limiting the validity and comparability of research on agile adoption.

This paper addresses this critical methodological gap by presenting the development of a novel instrument, the Manifesto Agreement Scale (MAS), and the systematic adaptation and validation of the Principle Agreement Scale (PAS), which is derived from the work of \citep{bishopUnderstandingPreferenceAgile2014}. We detail the process of item creation and selection, survey design, and rigorous psychometric validation analyses for both scales. Furthermore, we present a convergence and divergence analysis to empirically demonstrate that while the two scales are moderately correlated, they are not interchangeable and capture distinct dimensions of agile agreement.

The primary contribution of this work is a pair of publicly available instruments that have undergone a rigorous initial validation. By providing a reliable means to differentiate between agreement with agile values and practice, the MAS and PAS offer a foundation for a deeper understanding of agile team dynamics, the effectiveness of coaching interventions, and the multifaceted nature of the "person-agile fit." This study represents a critical first step, establishing a baseline for the cross-cultural and cross-sector validation required to develop a truly generalizable toolkit.

\section{Research Background}

\subsection{Conceptualizing Agile Agreement: Values vs. Practices} \label{sec:introvaluesvs_practices}

The development of agile methodologies was motivated by the necessity to effectively manage complex, rapidly changing projects~\citep{papadopoulosMovingTraditionalAgile2015}.
From traditional approaches, this represents a fundamental shift in perspective that emphasizes collaboration, flexibility, and continuous improvement~\citep{almeidaChallengesMigrationWaterfall2017}.
This paradigm was formalized with the publication of the Agile Manifesto in 2001, which aimed to "restore credibility to the word method" by delineating a novel framework for software engineering\citep{fowler2001agile,conboyConceptualFrameworkAgile2004}.
The Manifesto establishes four core values that prioritize individuals, working software, customer collaboration, and responsiveness to change over rigid processes, documentation, contracts, and plans.
A key nuance is that while the items on the right are acknowledged as having value, the items on the left are valued \textit{more} \citep{fowler2001agile}.
These values represent a high-level, abstract philosophy concerning the ideal state of software development.

Supporting this philosophy are the 12 Principles, which offer more concrete, actionable guidance.
Principles translate the abstract values into recommended operational behaviors and practices.
This creates a hierarchical structure from abstract values to concrete Principles, that is central to understanding agile adoption.

However, the distinction between agreement with the overarching philosophy (Values) and the day-to-day operational guidelines (Practices) is often conflated in both research and practice~\citep{gandomaniHowHumanAspects2014, kivAgileManifestoPractices2018, madiContentAnalysisAgile2011}.
Prior studies have shown a disconnect between practitioners' understanding of agile values and Principles.
For example, \citep{alzahraniInvestigatingAgileValues2024} found that prioritizing "Working software over comprehensive documentation" often leads to a tension between "more productivity over quality", highlighting a failure to balance this value with Principles like "Continuous attention to technical excellence".
This amalgamation is common, as practitioners often adopt practices without dedicating the necessary attention to the Manifesto that underpins them~\citep{kivAgileManifestoPractices2018}.

This conceptual ambiguity poses a significant challenge for measurement, as an individual may philosophically agree with a value (e.g., "Customer collaboration over contract negotiation") but disagree with the intensity of its practical application (e.g., daily collaboration).
This can lead to a superficial understanding of agile adoption, where teams may claim to adhere to the values while struggling with the implementation of the Principles~\citep{kivAgileManifestoPractices2018}.

\subsection{The Challenge of Measuring Agile Agreement}

Despite the acknowledged centrality of human factors in agile methodologies, the measurement of an individual's alignment with agile tenets remains underdeveloped.
A significant portion of the literature approaches this topic from a team-level perspective, employing instruments that assess constructs such as team climate~\citep{dutraUsingInstrumentAssess2023,dutraBuildingInstrumentAssess2023,dutraTACTInsTrumentAssess2022}, teamwork effectiveness~\citep{santosEvolutionTeamworkQuality2023,junkerAgileWorkPractices2023} or organizational maturity~\citep{looksStandardizedQuestionnaireMeasuring2021,grenProspectsQuantitativeMeasurement2015,tuncelSettingScopeNew2021}.
Other studies focus on team performance metrics rather than individual disposition~\citep{vishnubhotlaUnderstandingPerceivedRelevance2021,salamehPerformanceMeasurementFrameworkEvaluate2018,coelhoDesignModelApplication2020}.
Even when assessment occurs at the individual level, the focus is often on the \textit{perception} of practices and their implementation rather than personal \textit{agreement} with the underlying concepts. For instance, the Perceptive Agile Measurement (PAM) framework assesses an individual's perception of \textit{how} agile a team or project is functioning, not their personal \textit{endorsement} of that functioning~\citep{soPerceptiveAgileMeasurement2009}.
Similarly, other instruments designed to measure an "agile mindset" often treat agility as a monolithic concept. By design, these tools lack the necessary granularity to differentiate between the distinct conceptual levels of the Agile Manifesto, such as its core values, guiding principles, and specific practices~\citep{ozkanBackEssentialLiteratureBased2023,eilersWhyAgileMindset2021,stettinaFiveAgileFactors2011,florencioASAAgileSoftware2018}
The research most proximate to our objective is that of Bishop et al., which relates personality to a preference for the 12 agile Principles~\citep{bishopUnderstandingPreferenceAgile2014,bishopPersonalityTheoryPredictor2013,bishopAntecedentsPreferenceAgile2018}.
While valuable, this work does not explicitly distinguish agreement with the high-level values from agreement with the more concrete principles.\\

This review confirms a persistent gap, noted in prior systematic reviews, for objective and validated instruments to assess individual characteristics in agile contexts~\citep{dybaEmpiricalStudiesAgile2008a, gandomaniHowHumanAspects2014}.
Crucially, no existing instrument is explicitly designed to differentiate between an individual's agreement with the Manifesto's four abstract values and their practical agreement with its 12 concrete Principles.
This distinction is fundamental, as it would allow researchers to investigate critical questions: for instance, does values alignment necessarily translate to practical adherence? And how might different personality traits relate independently to these two distinct levels of agile agreement?
The development of such an instrument is, therefore, a necessary precondition for advancing a more nuanced, evidence-based understanding of the person-agile fit.

\section{Instrument Design and Methodology}

This section details the systematic methodology used for the construction and psychometric evaluation of the Manifesto Agreement Scale (MAS) and the Principle Agreement Scale (PAS). The process was designed to create objective measuring instruments capable of capturing the latent constructs associated with agile values and Principles. Adhering to established psychometric principles~\citep{dawis3ScaleConstruction2000,clarkConstructingValidityNew2019,maiti2021psychometric,el-denHowMeasureLatent2020,klineHandbookTestConstruction2015}, the primary goal was to maximize the construct validity of the scales—the fundamental consideration ensuring they precisely measure the theoretical constructs they were designed to capture. The methodology encompassed four key stages: (1) instrument design and item generation, (2) survey dissemination and data collection, (3) response filtering, and (4) a rigorous psychometric validation analysis to establish scale reliability and validity.

\subsection{Instrument Design and Item Generation}

A three-part questionnaire was developed in both English and French to capture the necessary constructs for our analysis. The development process was guided by established psychometric principles for creating objective measurement instruments. The primary goal was to ensure the construct validity of the scales, defined as the degree to which an instrument measures the theoretical construct it is intended to measure~\citep{dawis3ScaleConstruction2000,el-denHowMeasureLatent2020}. The concluding section of the survey gathered demographic data, as well as participants' self-assessed expertise in agile methodologies and their years of experience on agile teams. The first two parts, which form the core of this study, are detailed below.

\subsubsection{Manifesto Agreement Scale (MAS)}
To measure agreement with the abstract values of the Agile Manifesto, we designed a novel instrument from the ground up. This process began with a clear conceptualization of the target construct, which is the essential first step in scale development~\citep{el-denHowMeasureLatent2020,dawis3ScaleConstruction2000}. The complete survey instrument is accessible in~\ref{app:mas}. The Manifesto's four values are presented as preferential statements (e.g., "Individuals and interactions \textit{over} processes and tools"). To capture this nuance, we deconstructed each value into its two constituent components. From this conceptual model, an initial item pool was generated. For each of the resulting eight components (four "pro-agile," four "less-agile"), two distinct items were created to assess a respondent's agreement with applying that concept in practice. This use of multiple items is fundamental to covering the breadth of the construct and ensuring reliability. This resulted in a total of 16 items for the MAS. For example, the value "Working Software over comprehensive documentation" yielded four items: two assessing agreement with the importance of "Working Software" and two assessing agreement with the importance of "comprehensive documentation". All items were reviewed by a panel of 3 academic and 5 practitioner experts to ensure clarity, simple language, and lack of ambiguity~\citep{el-denHowMeasureLatent2020}. This review process involved conducting interviews with experts regarding each item to gather their feedback. This expert review serves two critical functions: (1) establishing face validity, or the degree to which the items appear to measure the construct, which is important for respondent motivation and cooperation ; and (2) establishing content validity, which is the "degree to which the items are relevant to and representative of the defined construct" as judged by content experts~\citep{maiti2021psychometric,klineHandbookTestConstruction2015}. Responses were collected on a four-point Likert-type scale  from "Totally Disagree" to "Completely Agree." While 5- or 7-point scales are common, an even number of response options was chosen. This design is psychometrically preferable for this construct as it avoids a "middle ground" or "central tendency" response bias and compels respondents to make a choice~\citep{dawis3ScaleConstruction2000}.

\subsubsection{Principle Agreement Scale (PAS)}
To measure agreement with the concrete practices derived from the 12 Agile Principles, we adapted and refined the instrument originally developed by \citep{bishopUnderstandingPreferenceAgile2014}. Adapting an existing, relevant instrument is a common and valid approach in scale development~\citep{el-denHowMeasureLatent2020}. Following their approach, we used the same consolidation of the 12 Principles into eight distinct conceptual groups. To ensure a manageable survey length while retaining the most representative items, we employed a systematic item selection process.
A panel of 8 experts, 3 from academic and 5 from industrial domain, ranked the items from the original 40-item questionnaire for each of the eight principle groups based on their representativeness. Representativeness was assessed through a survey in which participants were asked to rank items according to their significance relative to the targeted Principle. Additionally, items that were evaluated as very similar or duplicate were also noted by the experts. This reliance on expert judgment is a robust method for ensuring high content validity~\citep{dawis3ScaleConstruction2000}. The two most representative items for each group were selected from the initial item pool, resulting in a final 16-item scale. Each item retains the forced-choice, dichotomous format of the original, requiring respondents to choose between a statement aligned with agile practice and an alternative. The complete survey instrument is accessible in~\ref{app:pas}.

\subsection{Survey Dissemination and Participant Demographics}
The survey was disseminated online through university email networks in Belgium, the internal networks of major IT companies, and professional social media platforms such as LinkedIn. This strategy targeted active professionals in the software development field. We received 137 complete responses. The resulting sample was highly educated and professionally active, with 89\% of respondents holding at least a bachelor's degree and 85\% being employed. The predominant age range was 26-35 years (67\%), confirming that the survey reached its intended demographic.

A total of 137 participants began the survey, of which 113 complete responses were retained for the study, resulting in an 18\% drop-out rate. This sample size is considered adequate for initial psychometric validation, exceeding the minimal thresholds suggested for item analysis~\citep{dawis3ScaleConstruction2000,klineHandbookTestConstruction2015}.

\subsection{Psychometric Validation Methods}

A comprehensive psychometric evaluation was performed to establish the validity and reliability of the two newly developed scales, beginning with an assessment of reliability (the instrument's consistency) as an essential prerequisite for validity.

\subsubsection{Internal Consistency Analysis}
We assessed the internal consistency of each scale to ensure its items cohesively measure a single underlying construct. For the MAS, which uses a multi-point Likert scale, we calculated Cronbach's alpha. This coefficient is generally considered the most suitable and significant index for assessing the internal consistency reliability of rating scales~\citep{mokkinkCOSMINChecklistAssessing2010}. For the PAS, which uses a dichotomous scale, we used the Kuder-Richardson 20 (KR-20) formula. The KR-20 is the mathematically equivalent "special case of coefficient alpha" specifically designed for dichotomously scored items~\citep{feldtApproximateSamplingDistribution1965}.

\subsubsection{Split-Half Reliability Analysis}
To further assess reliability, we conducted a split-half analysis for both scales. This method also estimates internal consistency by correlating scores from two halves of the instrument. For the MAS, the 16 items were split into two halves based on the paired-item design, and the correlation between the two halves was calculated and adjusted using the Spearman-Brown correction. For the PAS, a similar split was performed.

\subsubsection{Construct Validity Analysis}
As part of establishing construct validity, we tested hypotheses about the scale's internal structure~\citep{dawis3ScaleConstruction2000,clarkConstructingValidityNew2019}. For the MAS, we assessed its structural construct validity by performing a specific split-half analysis that separated the eight "pro-agile" items from the eight "less-agile" items. A moderate negative correlation was hypothesized, as respondents who endorse pro-agile statements should logically tend to reject less-agile ones. This analysis tests the theoretical foundation upon which the items were generated.    

\subsubsection{Convergence and Divergence Analysis}

To determine whether the MAS and PAS measure related but distinct constructs (as hypothesized), we conducted an analysis of convergent and discriminant validity. This involved three analyses:
\begin{itemize}
    \item \textbf{Proportional Odds Logistic Regression}: To rigorously model the relationship between the \textit{Principle score} and the \textit{Manifest score}, a ordinal logistic model (also known as proportional odds logistic regression) is used. This approach is employed as our results derive from the combination of a sum of Likert scores and binary responses, resulting in ordinal, discrete outcomes. The implemented model of the \texttt{polr} function from the \texttt{MASS} package in R~\citep{venables2013modern} was used. Odds Ratios can be extracted from the model to quantify the effect size of the relationship between the Principle score and Manifest score categories.
    A fundamental prerequisite of the model is the \textbf{proportional odds assumption}. This assumption posits that the effect of the independent variable (Principle score) is uniform across all cumulative splits of the ordinal dependent variable. We will formally test this assumption using the Brant test~\citep{brant1990assessing}, implemented via the \texttt{brant} package. A non-significant omnibus $p$-value ($p > 0.05$) from the Brant test is required to confirm that this assumption is met, thereby validating the model's appropriateness.
    \item \textbf{Bland-Altman Plot:} To visually assess the agreement between the two scales. This graphical method plots the difference between the two scores on the $y$-axis against the mean of the two scores on the $x$-axis for each subject~\citep{altman1983measurement}. The resulting plot is used to evaluate two specific types of bias. (1) \textbf{Systematic Bias:} A mean difference that significantly diverges from 0, indicating that one scale systematically scores higher or lower than the other across all levels. (2) \textbf{Proportional Bias:} A difference between scores showing a trend as the mean score increases, suggesting that the agreement between the scales depends on the magnitude of the score~\citep{giavarinaUnderstandingBlandAltman2015}.
    \item \textbf{Intraclass Correlation Coefficient (ICC):} To provide a single, quantitative measure of reliability between the two scales. We will use the specific form \textbf{ICC(3,1)}, which corresponds to a two-way mixed-effects model assessing consistency. The resulting ICC coefficient, which ranges from 0 to 1, will be interpreted based on established guidelines (e.g., $<0.50$ = poor; $0.50-0.75$ = moderate; $0.75-0.90$ = good; $>0.90$ = excellent)~\citep{kooGuidelineSelectingReporting2016}.
\end{itemize}

\section{Results}

This section presents the results of the psychometric validation of the Manifesto Agreement Scale (MAS) and the Principle Agreement Scale (PAS). We first report the reliability and validity analyses for each scale individually and then present the convergence analysis that assesses their relationship.

\subsection{Psychometric Properties of the Scales}

\subsubsection{Manifesto Agreement Scale (MAS)}
The MAS demonstrated good internal consistency, achieving a Cronbach's alpha of $\alpha = 0.73$, which exceeds the standard threshold of 0.70 for acceptable reliability. This indicates that the items designed to measure agreement with the Manifesto's values form a cohesive scale. The split-half reliability analysis further supported this finding, yielding a Spearman-Brown corrected coefficient of $\rho_{xx'} = 0.80$, which indicates strong reliability.

Construct validity was assessed by correlating the eight "pro-agile" items with the eight "less-agile" items. This analysis yielded a moderate negative correlation ($\rho = -0.62$), confirming the hypothesis that respondents who endorse pro-agile values tend to reject the less-agile counterparts. This provides evidence that the scale is measuring the intended underlying construct.

\subsubsection{Principle Agreement Scale (PAS)}
Given the dichotomous nature of its items, the internal consistency of the PAS was evaluated using the Kuder-Richardson 20 formula. The scale achieved a reliability coefficient of $KR\text{-}20 = 0.71$, indicating satisfactory internal consistency. The split-half analysis confirmed this result with a Spearman-Brown corrected coefficient of $\rho_{xx'} = 0.83$, demonstrating excellent reliability.

\subsection{Convergence and Divergence of the MAS and PAS}

To evaluate whether the two scales measure related but distinct constructs, a convergence analysis was performed. Given the ordinal nature of the outcomes, a proportional odds logistic regression was implemented.The analysis yielded a statistically significant odds ratio of $1.66$ ($p < 0.05$). This result indicates that for a one-unit increase in the PAS score, the odds of a respondent being in a higher MAS score category (versus a lower one) increase by approximately $66\%$. While this confirms a positive association between the two scales, the magnitude of the effect suggests that the constructs are related but not strongly linked or interchangeable. Furthermore, the model's validity was confirmed, as the proportional odds assumption was met according to a non-significant Brant test (Omnibus $p = 0.73$). This result ensures the odds increase is stable and consistent across all categories of the MAS score.

\begin{figure}[h]
    \centering
    \includegraphics[width=.7\textwidth]{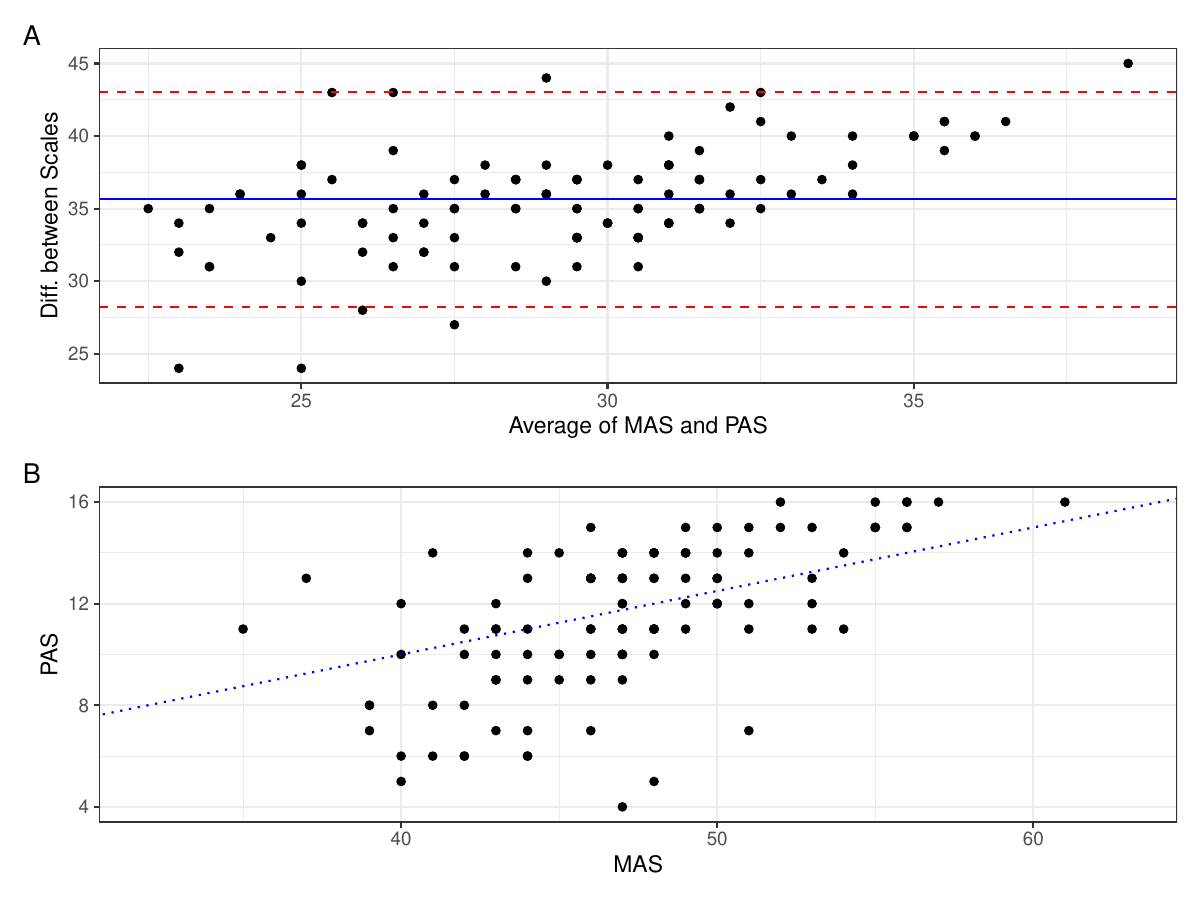}
    \caption{(A) Bland-Altman plot showing the difference between MAS and PAS scores against their average. The dashed lines represent the mean difference and the 95\% limits of agreement respectively. (B) Scatter plot of MAS vs. PAS scores, illustrating the relationship and variance at different score levels.}
    \label{fig:band_altman_ScatterPlot}
\end{figure}

A Bland-Altman plot was constructed to visualize the agreement between the two scales (see Figure~\ref{fig:band_altman_ScatterPlot}A). The plot revealed a significant proportional bias: the difference between the two scores increased as the average score increased. This indicates that while the scales show some agreement for individuals with low-to-moderate agile agreement, they diverge unilaterally for individuals with high agile agreement. This divergence is also visible in the scatter plot (Figure~\ref{fig:band_altman_ScatterPlot}B), which shows greater variance at higher score levels. The scatter plots further elucidate the proportional bias indicated in the Bland-Altman plot. While the dotted blue line represents a perfect alignment between the PAS and the MAS, the area below this line corresponds to individuals who align more closely with the MAS than with the PAS, whereas the area above the line indicates the opposite.

Finally, the Intraclass Correlation Coefficient (ICC) was calculated to provide a single measure of agreement. The analysis yielded a coefficient of $ICC(3,1) = 0.54$. According to established guidelines \citep{kooGuidelineSelectingReporting2016}, this value indicates only "moderate" agreement between the two scales.

These statistical findings directly address the conceptual framework established in section~\ref{sec:introvaluesvs_practices}, which posited that abstract \textit{Values} and concrete \textit{Practices} are distinct facets of agile agreement, each measured by MAS and PAS respectively. Taken together, these results provide empirical evidence that the MAS and PAS, while related, are not interchangeable, successfully capturing these two distinct facets: one related to abstract values and the other to a concrete practice.

\section{Discussion}

The primary contribution of this paper is the development and validation of two distinct instruments for measuring agile agreement: the novel Manifesto Agreement Scale (MAS) and the adapted Principle Agreement Scale (PAS). Our results demonstrate that both scales are reliable and valid measures. More importantly, the convergence analysis reveals that while the scales are moderately correlated, they are not interchangeable, providing empirical support for the conceptual distinction between agreeing with agile's abstract philosophy and its concrete practices.

\subsection{Interpreting the Methodological Findings}

The central finding of our validation is that measuring agreement with abstract values is different from measuring agreement with concrete practices. The moderate odds-ration ($1.66$) and moderate ICC value ($0.54$) indicate that the two constructs are related, as would be expected. However, the proportional bias revealed in the Bland-Altman plot is particularly insightful (figure~\ref{fig:band_altman_ScatterPlot}). It suggests that the distinction between values and practices becomes more pronounced for individuals with higher levels of agile agreement. This observation is particularly interesting for expert practitioners, such as agile experts and coaches, who depend on a nuanced understanding of these constructs.

This divergence may occur because it is relatively easy for anyone to agree with the high-level, abstract values of the Manifesto as they represent an idealized in a rather fuzzy level. In contrast, the Principles demand agreement with specific, and sometimes challenging, day-to-day behaviors. An individual might strongly endorse "Customer collaboration over contract negotiation" (a value) but find the idea of daily, intensive interaction with business stakeholders (a practice) to be impractical or undesirable (represented by the area below the line in the scatter plot, as shown in figure~\ref{fig:band_altman_ScatterPlot}B). Our instruments provide a means to capture this critical distinction, which has been overlooked by previous ad-hoc measurement approaches.

\subsection{Utility of the Instruments for Future Research}

The MAS and PAS provide a valuable toolkit for the empirical software engineering community, enabling more rigorous and nuanced research into the human factors of agile development.
These instruments open several avenues for future research. For instance, the scales can be utilized to quantitatively assess the effectiveness of agile coaching and training interventions.
Researchers could investigate whether a specific training program successfully enhances agreement on concrete practices, as measured by the PAS, even when abstract value agreement, measured by the MAS, was already high.
Furthermore, these instruments facilitate more granular research into team composition and dynamics, particularly the concept of "person-agile fit".
By distinguishing between value and practice agreement, researchers can explore more precise hypotheses, such as how specific personality traits correlate with each dimension of agreement, thereby contributing to a deeper understanding of team assembly.
Finally, in an organizational context, the scales could serve as a diagnostic tool to identify challenges in agile adoption.
A team exhibiting high MAS scores but low PAS scores would indicate that the primary challenge is not a lack of belief in the agile philosophy but rather friction with specific, mandated practices. Such a diagnosis would allow for more targeted and effective organizational interventions.

\subsection{Limitations}

It is important to acknowledge the limitations of this study. First, the scales were developed and validated using a sample drawn primarily from the Belgian IT sector. While the demographics are appropriate for this initial investigation, this single-country, relatively modest sample is insufficient to establish broad generalizability. Cultural contexts can significantly influence attitudes towards work practices and abstract values, representing a major confounding variable that has not been addressed in this study.
Further validation with more diverse, international samples is an essential next step to establish the instruments as general-purpose tools. Second, the MAS is an entirely novel instrument. As with any new instrument, ongoing research may identify opportunities to refine the wording of certain items to further improve clarity and reduce measurement noise. While the PAS is based on prior work, our specific adaptation and item selection also warrant further validation across different contexts.
Finally, this paper focused exclusively on the psychometric properties of the scales. The equally important question of how these measures relate to external criteria, such as individual performance or team effectiveness, remains a critical avenue for future research.

\section{Conclusion}

This study addressed a critical gap in empirical software engineering: the lack of validated instruments for measuring an individual's agreement with agile methodologies. We presented the development and validation of the novel Manifesto Agreement Scale (MAS) and the systematic adaptation of the Principle Agreement Scale (PAS).

Our analysis provides evidence that both instruments are reliable and valid. The central contribution of this work lies in the empirical demonstration that the MAS and PAS measure related but distinct constructs. This finding validates the conceptual distinction between agreement with agile's high-level, abstract values and its concrete, day-to-day practices—a nuance previously unaccounted for in measurement.

By providing these two publicly available instruments, validated within a specific demographic, this research offers a new foundation for more rigorous and granular investigations into the human factors of agile software development. The ability to differentiate between value and practice agreement will enable researchers to explore the complexities of the "person-agile fit," assess the effectiveness of coaching and training interventions, and diagnose adoption challenges with greater precision. This work serves as a critical baseline for the necessary cross-cultural and cross-sector validation studies that will ultimately advance a more evidence-based and generalizable understanding of how individuals and teams successfully engage with and adopt agile methodologies.

\section*{Disclosure of Interests}
The authors have no competing interests to declare that are relevant to the content of this article. AI tools were used for English language proofreading and flow improvement.

\appendix
\section{Manifesto Agreement Scale (MAS)} \label{app:mas}

The MAS consists of 16 items designed to measure agreement with the four core values of the Agile Manifesto. Each value is deconstructed into its two constituent components (one "pro-agile" (+), one "less-agile" (-)). Respondents rate their agreement with each statement on a four-point Likert scale ranging from "Totally Disagree" to "Completely Agree".

\begin{longtable}{@{}p{4.5cm} p{3.5cm} p{7.5cm}@{}}
    \caption{Items of the Manifesto Agreement Scale (MAS).} \label{tab:mvas_items}\\
    \toprule
    \textbf{Manifesto Value} & \textbf{Component} & \textbf{Item Statement} \\
    \midrule
    \endfirsthead
    \multicolumn{3}{c}%
    {{\bfseries \tablename\ \thetable{} -- continued from previous page}} \\
    \toprule
    \textbf{Manifesto Value} & \textbf{Component} & \textbf{Item Statement} \\
    \midrule
    \endhead
    \bottomrule
    \endfoot
    \multirow{4}{4.5cm}{Individuals and interactions \textit{over} processes and tools} & \multirow{2}{3.5cm}{Individuals and interactions (+)} & 1. Direct interactions with other team members are important to me. \\
    & & 2. A high-performing team must have the autonomy to organize its own work. \\
    \cmidrule{2-3}
    & \multirow{2}{3.5cm}{Processes and tools (-)} & 3. I work with difficulty without clearly established and complete procedures. \\
    & & 4. I prefer using a task management tool over talking directly to my colleagues to know the progress. \\
    \midrule
    \multirow{4}{4.5cm}{Working software \textit{over} comprehensive documentation} & \multirow{2}{3.5cm}{Working software (+)} & 5. Aiming for a working solution above all else is essential for me. \\
    & & 6. The main measure of a project's progress is the availability of new features. \\
    \cmidrule{2-3}
    & \multirow{2}{3.5cm}{Comprehensive documentation (-)} & 7. An important part of the work must focus on writing complete and exhaustive documentation. \\
    & & 8. Every project must first be completely detailed before starting any development. \\
    \midrule
    \multirow{4}{4.5cm}{Customer collaboration \textit{over} contract negotiation} & \multirow{2}{3.5cm}{Customer collaboration (+)} & 9. It is crucial to continuously integrate client feedback throughout the project. \\
    & & 10. Interacting regularly with the client allows me to be sure I'm going in the right direction. \\
    \cmidrule{2-3}
    & \multirow{2}{3.5cm}{Contract negotiation (-)} & 11. Interactions with the client only happen before and after the completion of a project. \\
    & & 12. All decisions made at the start of the project are firm and unchangeable. \\
    \midrule
    \multirow{4}{4.5cm}{Responding to change \textit{over} following a plan} & \multirow{2}{3.5cm}{Responding to change (+)} & 13. I see changes requested by the client, even late in the project, as opportunities to improve the final product. \\
    & & 14. I have no problem reworking already developed functionalities to better meet the client's needs. \\
    \cmidrule{2-3}
    & \multirow{2}{3.5cm}{Following a plan (-)} & 15. It is important for me not to deviate from the work plan validated at the start of the project. \\
    & & 16. A document that clearly describes all the steps of the project is essential for me. \\
\end{longtable}

\section{Principle Agreement Scale (PAS)} \label{app:pas}
The PAS consists of 16 item pairs designed to measure agreement with practices derived from the 12 Agile Principles. The Principles were consolidated into eight conceptual groups. For each item, respondents must make a forced-choice selection between two statements, one of which represents a pro-agile practice. The pro-agile choice is marked with (*) in the table below.

\begin{longtable}{@{}p{2cm} p{5cm} p{8cm}@{}}
    \caption{Items of the Principle Agreement Scale (PAS).} \label{tab:ppas_items_full}\\
    \toprule
    \textbf{Principle Group} & \textbf{Question Stem} & \textbf{Choices} \\
    \midrule
    \endfirsthead
    \multicolumn{3}{c}%
    {{\bfseries \tablename\ \thetable{} -- continued from previous page}} \\
    \toprule
    \textbf{Principle Group} & \textbf{Question Stem} & \textbf{Choices} \\
    \midrule
    \endhead
    \bottomrule
    \endfoot
    \multirow{4}{*}{1 \& 3} & \multirow{2}{5cm}{I believe customer satisfaction is best achieved by...} & A. Using Gantt charts to demonstrate how we meet their requirements. \\
    & & B. Delivering a new version of the product each month to offer them new working functionalities. (*) \\
    \cmidrule{2-3}
    & \multirow{2}{5cm}{I enjoy...} & A. Working in longer phases and delivering a finished product. \\
    & & B. Frequently delivering smaller increments of the solution, even if not all functionalities are implemented. (*) \\
    \midrule
    \multirow{4}{*}{2} & \multirow{2}{5cm}{I believe changes to the requirements mean...} & A. that the client has a competitive advantage in the market and that they should be welcomed favorably. (*) \\
    & & B. that it involves significant extra work and should be avoided. \\
    \cmidrule{2-3}
    & \multirow{2}{5cm}{I like it when...} & A. requirements changes occur at any time so that we get what is needed. (*) \\
    & & B. user requests are approved and finalized before working on the product. \\
    \midrule
    \multirow{4}{*}{4} & \multirow{2}{5cm}{I prefer...} & A. Working with the users of the solution on a daily basis. (*) \\
    & & B. Working with written specifications and documents. \\
    \cmidrule{2-3}
    & \multirow{2}{5cm}{I like it when business users...} & A. engage with the entire development team. (*) \\
    & & B. interact with competent analysts to define the requirements. \\
    \midrule
    \multirow{4}{*}{6} & \multirow{2}{5cm}{I prefer to get my information...} & A. by using documents and diagrams. \\
    & & B. through face-to-face communication. (*) \\
    \cmidrule{2-3}
    & \multirow{2}{5cm}{Given the choice...} & A. I would prefer to be personally involved, get the user's needs in writing, and then create the product to meet their needs. \\
    & & B. I would prefer to interact daily with the people who will use the software to ensure I know what they want. (*) \\
    \midrule
    \multirow{4}{*}{7} & \multirow{2}{5cm}{I like...} & A. to use the delivered product as a measure of progress made. (*) \\
    & & B. to complete the tasks assigned to me and use that to measure my progress on the work plan. \\
    \cmidrule{2-3}
    & \multirow{2}{5cm}{The best measure of progress is...} & A. the percentage of tasks completed on the project schedule. \\
    & & B. the number of functionalities (or achievements) validated by the client. (*) \\
    \midrule
    \multirow{4}{*}{8} & \multirow{2}{5cm}{I'd rather...} & A. work in waves with waiting periods and higher-intensity periods. \\
    & & B. work at a constant pace over the long term. (*) \\
    \cmidrule{2-3}
    & \multirow{2}{5cm}{Given the pace of business today I believe...} & A. that it is essential for the workload to remain sustainable to ensure a long-term vision. (*) \\
    & & B. that it is inevitable that there will be peaks in workload to complete projects and that teams must work harder during certain periods. \\
    \midrule
    \multirow{4}{*}{11} & \multirow{2}{5cm}{I believe architecture, requirements and design should be...} & A. determined by the team based on its internal cooperation and self-organization. (*) \\
    & & B. assigned by the project manager to the appropriate competent people within a group. \\
    \cmidrule{2-3}
    & \multirow{2}{5cm}{I believe teams work better when...} & A. a project manager directs the work according to the roles and missions of the project. \\
    & & B. the team self-organizes by making collective decisions on how the work should be done. (*) \\
    \midrule
    \multirow{4}{*}{12} & \multirow{2}{5cm}{It is important to...} & A. follow the project manager's recommendations on group effectiveness. \\
    & & B. reflect together as a group on how to become more effective. (*) \\
    \cmidrule{2-3}
    & \multirow{2}{5cm}{I think to become more effective...} & A. the team should regularly reflect on its work practices and its behavior in the relevant areas. (*) \\
    & & B. at the end of the project's completion, the leadership should regularly examine the team's practices and make suggestions for improvement. \\
\end{longtable}

\bibliographystyle{unsrtnat}
\bibliography{references}  

@book{venables2013modern,
  title = {Modern Applied Statistics with {{S}}},
  author = {Venables, William N and Ripley, Brian D},
  year = {2013},
  publisher = {Springer Science \& Business Media}
}

@article{brant1990assessing,
  title = {Assessing Proportionality in the Proportional Odds Model for Ordinal Logistic Regression},
  author = {Brant, Rollin},
  year = {1990},
  journaltitle = {Biometrics. Journal of the International Biometric Society},
  shortjournal = {Biometrics},
  pages = {1171--1178},
  publisher = {JSTOR}
}

@article{giavarinaUnderstandingBlandAltman2015,
  title = {Understanding {{Bland Altman}} Analysis},
  author = {Giavarina, Davide},
  year = {2015},
  journaltitle = {Biochemia Medica},
  shortjournal = {Biochem Med},
  volume = {25},
  number = {2},
  pages = {141--151},
  publisher = {{Croatian Society of Medical Biochemistry and Laboratory Medicine}},
  doi = {10.11613/BM.2015.015},
  url = {https://www.biochemia-medica.com/en/journal/25/10.11613/BM.2015.015},
  urldate = {2025-11-06},
  langid = {english}
}

@article{altman1983measurement,
  title = {Measurement in Medicine: The Analysis of Method Comparison Studies},
  author = {Altman, Douglas G and Bland, J Martin},
  year = {1983},
  journaltitle = {Journal of the Royal Statistical Society Series D: The Statistician},
  volume = {32},
  number = {3},
  pages = {307--317},
  publisher = {Oxford University Press}
}

@article{clarkConstructingValidityNew2019,
  title = {Constructing Validity: {{New}} Developments in Creating Objective Measuring Instruments},
  shorttitle = {Constructing Validity},
  author = {Clark, Lee Anna and Watson, David},
  date = {2019},
  year = {2019},
  journaltitle = {Psychological Assessment},
  volume = {31},
  number = {12},
  pages = {1412--1427},
  publisher = {American Psychological Association},
  location = {US},
  issn = {1939-134X},
  doi = {10.1037/pas0000626}
}

@incollection{dawis3ScaleConstruction2000,
  title = {3 - {{Scale Construction}} and {{Psychometric Considerations}}},
  booktitle = {Handbook of {{Applied Multivariate Statistics}} and {{Mathematical Modeling}}},
  author = {Dawis, Rene V.},
  editor = {Tinsley, Howard E. A. and Brown, Steven D.},
  date = {2000-01-01},
  year={2000},
  pages = {65--94},
  publisher = {Academic Press},
  location = {San Diego},
  doi = {10.1016/B978-012691360-6/50004-5},
  url = {https://www.sciencedirect.com/science/article/pii/B9780126913606500045},
  urldate = {2025-10-30},
  isbn = {978-0-12-691360-6}
}

@article{el-denHowMeasureLatent2020,
  title = {How to Measure a Latent Construct: {{Psychometric}} Principles for the Development and Validation of Measurement Instruments},
  shorttitle = {How to Measure a Latent Construct},
  author = {El-Den, Sarira and Schneider, Carl and Mirzaei, Ardalan and Carter, Stephen},
  date = {2020-08-01},
  year = {2020},
  journaltitle = {International Journal of Pharmacy Practice},
  shortjournal = {Int J Pharm Pract},
  volume = {28},
  number = {4},
  pages = {326--336},
  issn = {0961-7671},
  doi = {10.1111/ijpp.12600},
  url = {https://doi.org/10.1111/ijpp.12600},
  urldate = {2025-10-30}
}

@article{feldtApproximateSamplingDistribution1965,
  title = {The Approximate Sampling Distribution of {{Kuder-Richardson}} Reliability Coefficient Twenty},
  author = {Feldt, Leonard S.},
  date = {1965-09-01},
  year = {1965},
  journaltitle = {Psychometrika},
  shortjournal = {Psychometrika},
  volume = {30},
  number = {3},
  pages = {357--370},
  issn = {1860-0980},
  doi = {10.1007/BF02289499},
  url = {https://doi.org/10.1007/BF02289499},
  urldate = {2025-10-30},
  langid = {english}
}

@book{klineHandbookTestConstruction2015,
  title = {A {{Handbook}} of {{Test Construction}} ({{Psychology Revivals}}): {{Introduction}} to {{Psychometric Design}}},
  shorttitle = {A {{Handbook}} of {{Test Construction}} ({{Psychology Revivals}})},
  author = {Kline, Paul},
  date = {2015-06-03},
  year = {2015},
  publisher = {Routledge},
  location = {London},
  doi = {10.4324/9781315695990},
  isbn = {978-1-315-69599-0},
  pagetotal = {274}
}

@article{maiti2021psychometric,
  title = {Psychometric Scale Construction Techniques: Basics to Advances},
  author = {Maiti, Sanjit and Garai, Sanchita and Mohammad, Asif and family=Kadian, given=SK, given-i=SK},
  year = {2021},
  journaltitle = {Dairy Extension Division, ICAR-National Dairy Research Institute, Karnal, Haryana}
}

@article{mokkinkCOSMINChecklistAssessing2010,
  title = {The {{COSMIN}} Checklist for Assessing the Methodological Quality of Studies on Measurement Properties of Health Status Measurement Instruments: An International {{Delphi}} Study},
  shorttitle = {The {{COSMIN}} Checklist for Assessing the Methodological Quality of Studies on Measurement Properties of Health Status Measurement Instruments},
  author = {Mokkink, Lidwine B. and Terwee, Caroline B. and Patrick, Donald L. and Alonso, Jordi and Stratford, Paul W. and Knol, Dirk L. and Bouter, Lex M. and de Vet, Henrica C. W.},
  date = {2010-05-01},
  year = {2010},
  journaltitle = {Quality of Life Research},
  shortjournal = {Qual Life Res},
  volume = {19},
  number = {4},
  pages = {539--549},
  issn = {1573-2649},
  doi = {10.1007/s11136-010-9606-8},
  url = {https://doi.org/10.1007/s11136-010-9606-8},
  urldate = {2025-10-30},
  langid = {english}
}

@article{almeidaChallengesMigrationWaterfall2017,
  title = {Challenges in {{Migration}} from {{Waterfall}} to {{Agile Environments}}},
  author = {Almeida, Fernando},
  year = {2017},
  journal = {World Journal of Computer Application and Technology(CEASE PUBLICATION)},
  volume = {5},
  number = {3},
  pages = {39--49},
  publisher = {Horizon Research Publishing},
  doi = {10.13189/wjcat.2017.050302},
  urldate = {2025-05-26},
  copyright = {https://creativecommons.org/licenses/by/4.0/},
  langid = {english}
}

@article{alzahraniInvestigatingAgileValues2024,
  title = {Investigating {{Agile Values}} and {{Principles}} in {{Real Practices}}},
  author = {Alzahrani, Abdullah A. H.},
  year = {2024},
  month = jan,
  journal = {International Journal of Advanced Computer Science and Applications (IJACSA)},
  volume = {15},
  number = {1},
  publisher = {{The Science and Information (SAI) Organization Limited}},
  issn = {2156-5570},
  doi = {10.14569/IJACSA.2024.0150128},
  urldate = {2025-10-03},
  langid = {english}
}

@article{bishopAntecedentsPreferenceAgile2018,
  title = {Antecedents of {{Preference}} for {{Agile Methods}}: {{A Project Manager Perspective}}},
  shorttitle = {Antecedents of {{Preference}} for {{Agile Methods}}},
  author = {Bishop, David and Rowland, Pam and Noteboom, Cherie},
  year = {2018},
  month = jan,
  journal = {Hawaii International Conference on System Sciences 2018 (HICSS-51)}
}

@article{bishopPersonalityTheoryPredictor2013,
  title = {Personality {{Theory}} as a {{Predictor}} for {{Agile Preference}}},
  author = {Bishop, Dave},
  year = {2013},
  month = may,
  journal = {Faculty Research \& Publications}
}

@inproceedings{bishopUnderstandingPreferenceAgile2014,
  title = {Toward an {{Understanding}} of {{Preference}} for {{Agile Software Development Methods}} from a {{Personality Theory Perspective}}},
  booktitle = {2014 47th {{Hawaii International Conference}} on {{System Sciences}}},
  author = {Bishop, David and Deokar, Amit},
  year = {2014},
  month = jan,
  pages = {4749--4758},
  issn = {1530-1605},
  doi = {10.1109/HICSS.2014.583}
}

@inproceedings{coelhoDesignModelApplication2020,
  title = {Design of {{Model}} and {{Application Development}} for {{Agile Practices Assessment}}},
  booktitle = {2020 15th {{Iberian Conference}} on {{Information Systems}} and {{Technologies}} ({{CISTI}})},
  author = {Coelho, Isabel and Ventura, Paula and Reis, Leonilde},
  year = {2020},
  month = jun,
  pages = {1--5},
  issn = {2166-0727},
  doi = {10.23919/CISTI49556.2020.9140849},
  urldate = {2025-10-03}
}

@inproceedings{conboyConceptualFrameworkAgile2004,
  title = {Toward a {{Conceptual Framework}} of {{Agile Methods}}},
  booktitle = {Extreme {{Programming}} and {{Agile Methods}} - {{XP}}/{{Agile Universe}} 2004},
  author = {Conboy, Kieran and Fitzgerald, Brian},
  editor = {Zannier, Carmen and Erdogmus, Hakan and Lindstrom, Lowell},
  year = {2004},
  pages = {105--116},
  publisher = {Springer},
  address = {Berlin, Heidelberg},
  doi = {10.1007/978-3-540-27777-4_11},
  isbn = {978-3-540-27777-4},
  langid = {english}
}

@article{daraojimbaCOMPREHENSIVEREVIEWAGILE2024,
  title = {{{COMPREHENSIVE REVIEW OF AGILE METHODOLOGIES IN PROJECT MANAGEMENT}}},
  author = {Daraojimba, Emmanuel Chibuike and Nwasike, Chinedu Nnamdi and Adegbite, Abimbola Oluwatoyin and Ezeigweneme, Chinedu Alex and Gidiagba, Joachim Osheyor},
  year = {2024},
  month = jan,
  journal = {Computer Science \& IT Research Journal},
  volume = {5},
  number = {1},
  pages = {190--218},
  issn = {2709-0051},
  doi = {10.51594/csitrj.v5i1.717},
  urldate = {2025-06-05},
  copyright = {Copyright (c) 2024 Emmanuel Chibuike Daraojimba, Chinedu Nnamdi Nwasike, Abimbola Oluwatoyin Adegbite, Chinedu Alex Ezeigweneme, Joachim Osheyor Gidiagba},
  langid = {english}
}

@inproceedings{dutraBuildingInstrumentAssess2023,
  title = {On {{Building}} an {{Instrument}} to {{Assess}} the {{Organizational Climate}} of {{Agile Software Development Teams}}},
  booktitle = {Proceedings of the {{XXII Brazilian Symposium}} on {{Software Quality}}},
  author = {Dutra, Eliezer and Diirr, Bruna and Santos, Gleison},
  year = {2023},
  month = dec,
  series = {{{SBQS}} '23},
  pages = {342--351},
  publisher = {Association for Computing Machinery},
  address = {New York, NY, USA},
  doi = {10.1145/3629479.3629492},
  urldate = {2025-10-03},
  isbn = {979-8-4007-0786-5}
}

@article{dutraTACTInsTrumentAssess2022,
  title = {{{TACT}}: {{An insTrument}} to {{Assess}} the Organizational {{ClimaTe}} of Agile Teams - {{A Preliminary Study}}},
  shorttitle = {{{TACT}}},
  author = {Dutra, Eliezer and Lima, Patr{\'i}cia and Cerdeiral, Cristina and Diirr, Bruna and Santos, Gleison},
  year = {2022},
  month = jan,
  journal = {Journal of Software Engineering Research and Development},
  volume = {10},
  pages = {1:1-1:21},
  issn = {2195-1721},
  doi = {10.5753/jserd.2021.1973},
  urldate = {2025-10-03},
  copyright = {Copyright (c) 2022 Eliezer Goncalves, Patr{\'i}cia Lima, Cristina Cerdeiral, Bruna Diirr, Gleison Santos},
  langid = {english}
}

@article{dutraUsingInstrumentAssess2023,
  title = {Using an {{Instrument}} to {{Assess Trust}}, {{Knowledge}}, {{Learning}}, and {{Motivation}} of {{Agile Teams}}},
  author = {Dutra, Eliezer and Cerdeiral, Cristina and Lima, Patr{\'i}cia and Escalfoni, Rafael and Diirr, Bruna and Santos, Gleison},
  year = {2023},
  month = may,
  journal = {iSys - Brazilian Journal of Information Systems},
  volume = {16},
  number = {1},
  pages = {7:1-7:32},
  issn = {1984-2902},
  doi = {10.5753/isys.2023.3005},
  urldate = {2025-10-03},
  copyright = {Copyright (c) 2023 iSys - Brazilian Journal of Information Systems},
  langid = {english}
}

@article{dybaEmpiricalStudiesAgile2008a,
  title = {Empirical Studies of Agile Software Development: {{A}} Systematic Review},
  shorttitle = {Empirical Studies of Agile Software Development},
  author = {Dyb{\aa}, Tore and Dings{\o}yr, Torgeir},
  year = {2008},
  month = aug,
  journal = {Information and Software Technology},
  volume = {50},
  number = {9},
  pages = {833--859},
  issn = {0950-5849},
  doi = {10.1016/j.infsof.2008.01.006},
  urldate = {2024-10-07}
}

@article{eilersWhyAgileMindset2021,
  title = {Why the {{Agile Mindset Matters}}},
  author = {Eilers, Karen and Simmert, Benedikt and Peters, Christoph and Leimeister, Jan Marco},
  year = {2021},
  month = aug,
  journal = {Academy of Management Proceedings},
  volume = {2021},
  number = {1},
  pages = {13110},
  publisher = {Academy of Management},
  issn = {0065-0668},
  doi = {10.5465/AMBPP.2021.13110abstract},
  urldate = {2025-07-01}
}

@inproceedings{florencioASAAgileSoftware2018,
  title = {{{ASA}}: {{Agile Software Development Self-assessment Method}}},
  shorttitle = {{{ASA}}},
  booktitle = {Agile {{Methods}}},
  author = {Flor{\^e}ncio, Matheus and Sambinelli, Fernando and Francisco Borges, Marcos Augusto},
  editor = {dos Santos, Viviane Almeida and Pinto, Gustavo Henrique Lima and Serra Seca Neto, Adolfo Gustavo},
  year = {2018},
  pages = {21--30},
  publisher = {Springer International Publishing},
  address = {Cham},
  doi = {10.1007/978-3-319-73673-0_2},
  isbn = {978-3-319-73673-0},
  langid = {english}
}

@article{fowler2001agile,
  title = {The Agile Manifesto},
  author = {Fowler, Martin and Highsmith, Jim},
  year = {2001},
  month = aug,
  journal = {Software Development}
}

@article{gandomaniHowHumanAspects2014,
  title = {How {{Human Aspects Impress Agile Software Development Transition}} and {{Adoption}}},
  author = {Gandomani, Taghi Javdani and Zulzalil, Hazura and Ghani, Abdul Azim Abdul and {Abu Bakar Sultan} and Sharif, Khaironi Yatim},
  year = {2014},
  month = jan,
  journal = {International Journal of Software Engineering and its Applications},
  volume = {8},
  number = {1},
  pages = {129--148},
  doi = {10.14257/ijseia.2014.8.1.12}
}

@article{grenProspectsQuantitativeMeasurement2015,
  title = {The Prospects of a Quantitative Measurement of Agility: {{A}} Validation Study on an Agile Maturity Model},
  shorttitle = {The Prospects of a Quantitative Measurement of Agility},
  author = {Gren, Lucas and Torkar, Richard and Feldt, Robert},
  year = {2015},
  month = sep,
  journal = {Journal of Systems and Software},
  volume = {107},
  pages = {38--49},
  issn = {0164-1212},
  doi = {10.1016/j.jss.2015.05.008},
  urldate = {2025-10-03}
}

@article{junkerAgileWorkPractices2023,
  title = {Agile Work Practices: Measurement and Mechanisms},
  shorttitle = {Agile Work Practices},
  author = {Junker, Tom L. and Bakker, Arnold B. and Derks, Daantje and Molenaar, Dylan},
  year = {2023},
  month = jan,
  journal = {European Journal of Work and Organizational Psychology},
  volume = {32},
  number = {1},
  pages = {1--22},
  publisher = {Routledge},
  issn = {1359-432X},
  doi = {10.1080/1359432X.2022.2096439},
  urldate = {2025-10-03}
}

@inproceedings{kivAgileManifestoPractices2018,
  title = {Agile {{Manifesto}} and {{Practices Selection}} for {{Tailoring Software Development}}: {{A Systematic Literature Review}}},
  shorttitle = {Agile {{Manifesto}} and {{Practices Selection}} for {{Tailoring Software Development}}},
  booktitle = {Product-{{Focused Software Process Improvement}}},
  author = {Kiv, Soreangsey and Heng, Samedi and Kolp, Manuel and Wautelet, Yves},
  editor = {Kuhrmann, Marco and Schneider, Kurt and Pfahl, Dietmar and Amasaki, Sousuke and Ciolkowski, Marcus and Hebig, Regina and Tell, Paolo and Kl{\"u}nder, Jil and K{\"u}pper, Steffen},
  year = {2018},
  pages = {12--30},
  publisher = {Springer International Publishing},
  address = {Cham},
  doi = {10.1007/978-3-030-03673-7_2},
  isbn = {978-3-030-03673-7},
  langid = {english}
}

@article{kooGuidelineSelectingReporting2016,
  title = {A {{Guideline}} of {{Selecting}} and {{Reporting Intraclass Correlation Coefficients}} for {{Reliability Research}}},
  author = {Koo, Terry K. and Li, Mae Y.},
  year = {2016},
  month = jun,
  journal = {Journal of Chiropractic Medicine},
  volume = {15},
  number = {2},
  pages = {155--163},
  issn = {1556-3707},
  doi = {10.1016/j.jcm.2016.02.012},
  urldate = {2025-07-11},
  pmcid = {PMC4913118},
  pmid = {27330520}
}

@inproceedings{looksStandardizedQuestionnaireMeasuring2021,
  title = {Towards a {{Standardized Questionnaire}} for {{Measuring Agility}} at {{Team Level}}},
  booktitle = {Agile {{Processes}} in {{Software Engineering}} and {{Extreme Programming}}},
  author = {Looks, Hanna and Fangmann, Jannik and Thomaschewski, J{\"o}rg and Escalona, Mar{\'i}a-Jos{\'e} and Sch{\"o}n, Eva-Maria},
  editor = {Gregory, Peggy and Lassenius, Casper and Wang, Xiaofeng and Kruchten, Philippe},
  year = {2021},
  pages = {71--85},
  publisher = {Springer International Publishing},
  address = {Cham},
  doi = {10.1007/978-3-030-78098-2_5},
  isbn = {978-3-030-78098-2},
  langid = {english}
}

@inproceedings{madiContentAnalysisAgile2011,
  title = {Content Analysis on Agile Values: {{A}} Perception from Software Practitioners},
  shorttitle = {Content Analysis on Agile Values},
  booktitle = {2011 {{Malaysian Conference}} in {{Software Engineering}}},
  author = {Madi, Tamer and Dahalin, Zulkhairi and Baharom, Fauziah},
  year = {2011},
  month = dec,
  pages = {423--428},
  doi = {10.1109/MySEC.2011.6140710},
  urldate = {2025-10-03}
}

@inproceedings{ozkanBackEssentialLiteratureBased2023,
  title = {Back to the {{Essential}}: {{A Literature-Based Review}} on {{Agile Mindset}}},
  shorttitle = {Back to the {{Essential}}},
  booktitle = {Annals of {{Computer Science}} and {{Information Systems}}},
  author = {Ozkan, Necmettin and Eilers, Karen and G{\"o}k, Mehmet {\c S}ahin},
  year = {2023},
  volume = {35},
  pages = {201--211},
  issn = {2300-5963},
  urldate = {2025-07-01},
  isbn = {978-83-967447-8-4},
  langid = {english}
}

@article{papadopoulosMovingTraditionalAgile2015,
  title = {Moving from {{Traditional}} to {{Agile Software Development Methodologies Also}} on {{Large}}, {{Distributed Projects}}.},
  author = {Papadopoulos, Georgios},
  year = {2015},
  month = feb,
  journal = {Procedia - Social and Behavioral Sciences},
  series = {Proceedings of the 3rd {{International Conference}} on {{Strategic Innovative Marketing}} ({{IC-SIM}} 2014)},
  volume = {175},
  pages = {455--463},
  issn = {1877-0428},
  doi = {10.1016/j.sbspro.2015.01.1223},
  urldate = {2025-05-26}
}

@article{salamehPerformanceMeasurementFrameworkEvaluate2018,
  title = {Performance-{{Measurement Framework}} to {{Evaluate Software Engineers}} for {{Agile Software-Development Methodology}}},
  author = {Salameh, Hanadi},
  year = {2018},
  month = mar,
  journal = {European Journal of International Management},
  volume = {Vol.7}
}

@inproceedings{santosEvolutionTeamworkQuality2023,
  title = {Evolution of {{Teamwork Quality Instruments}} in {{Agile Software Development}}: {{A Systematic Literature Review}}},
  shorttitle = {Evolution of {{Teamwork Quality Instruments}} in {{Agile Software Development}}},
  booktitle = {Proceedings of the {{XXXVII Brazilian Symposium}} on {{Software Engineering}}},
  author = {dos Santos, Ramon N{\'o}brega and Cunha, Felipe Oliveira Miranda and Rique, Thiago Pereira and Perkusich, Mirko and Neto, Ademar Fran{\c c}a de Sousa and Albuquerque, Danyllo Wagner and Almeida, Hyggo and Perkusich, Angelo},
  year = {2023},
  month = sep,
  series = {{{SBES}} '23},
  pages = {216--225},
  publisher = {Association for Computing Machinery},
  address = {New York, NY, USA},
  doi = {10.1145/3613372.3613404},
  urldate = {2025-10-03},
  isbn = {979-8-4007-0787-2}
}

@inproceedings{soPerceptiveAgileMeasurement2009,
  title = {Perceptive {{Agile Measurement}}: {{New Instruments}} for {{Quantitative Studies}} in the {{Pursuit}} of the {{Social-Psychological Effect}} of {{Agile Practices}}},
  shorttitle = {Perceptive {{Agile Measurement}}},
  booktitle = {Agile {{Processes}} in {{Software Engineering}} and {{Extreme Programming}}},
  author = {So, Chaehan and Scholl, Wolfgang},
  editor = {Abrahamsson, Pekka and Marchesi, Michele and Maurer, Frank},
  year = {2009},
  pages = {83--93},
  publisher = {Springer},
  address = {Berlin, Heidelberg},
  doi = {10.1007/978-3-642-01853-4_11},
  isbn = {978-3-642-01853-4},
  langid = {english}
}

@inproceedings{stettinaFiveAgileFactors2011,
  title = {Five {{Agile Factors}}: {{Helping Self-management}} to {{Self-reflect}}},
  shorttitle = {Five {{Agile Factors}}},
  booktitle = {Systems, {{Software}} and {{Service Process Improvement}}},
  author = {Stettina, Christoph J. and Heijstek, Werner},
  editor = {O`Connor, Rory V. and {Pries-Heje}, Jan and Messnarz, Richard},
  year = {2011},
  pages = {84--96},
  publisher = {Springer},
  address = {Berlin, Heidelberg},
  doi = {10.1007/978-3-642-22206-1_8},
  isbn = {978-3-642-22206-1},
  langid = {english}
}

@inproceedings{tuncelSettingScopeNew2021,
  title = {Setting the {{Scope}} for a {{New Agile Assessment Model}}: {{Results}} of an {{Empirical Study}}},
  shorttitle = {Setting the {{Scope}} for a {{New Agile Assessment Model}}},
  booktitle = {Agile {{Processes}} in {{Software Engineering}} and {{Extreme Programming}}},
  author = {Tuncel, Doruk and K{\"o}rner, Christian and Pl{\"o}sch, Reinhold},
  editor = {Gregory, Peggy and Lassenius, Casper and Wang, Xiaofeng and Kruchten, Philippe},
  year = {2021},
  pages = {55--70},
  publisher = {Springer International Publishing},
  address = {Cham},
  doi = {10.1007/978-3-030-78098-2_4},
  isbn = {978-3-030-78098-2},
  langid = {english}
}

@article{vishnubhotlaUnderstandingPerceivedRelevance2021,
  title = {Understanding the Perceived Relevance of Capability Measures: {{A}} Survey of {{Agile Software Development}} Practitioners},
  shorttitle = {Understanding the Perceived Relevance of Capability Measures},
  author = {Vishnubhotla, Sai Datta and Mendes, Emilia and Lundberg, Lars},
  year = {2021},
  month = oct,
  journal = {Journal of Systems and Software},
  volume = {180},
  pages = {111013},
  issn = {0164-1212},
  doi = {10.1016/j.jss.2021.111013},
  urldate = {2025-10-03}
}






\end{document}